\documentclass[useAMS,usenatbib]{mn2e}
\usepackage[utf8]{inputenc}
\usepackage {graphicx,color}
\usepackage{xcolor}

\newcommand{\msun}{\mbox{${\rm M_{\sun}}$}}
\title[Impact of thermal diffusion]{Impact of thermal diffusion and
  other abundance anomalies on cosmological uses of galaxy clusters} 
\author[P. Medvedev et al.]{P. Medvedev$^{1}$\thanks{E-mail:
tomedvedev@iki.rssi.ru}, M.~Gilfanov$^{1,2}$, S. Sazonov$^{1,3}$,
  P. Shtykovskiy$^{1}$\\ 
$^{1}$Space Research Institute, Russian Academy of Sciences,
  Profsoyuznaya 84/32, 117997 Moscow, Russia\\ 
$^{2}$Max-Planck-Institut für Astrophysik, Karl-Schwarzschild-Str. 1,
  D-85741 Garching bei München, Germany\\ 
$^{3}$Moscow Institute of Physics and Technology, 9 Institutsky
  per., 141700 Dolgoprudny, Moscow Region, Russia} 
\begin{document}

\date{Aug. 2013}

\pagerange{\pageref{firstpage}--\pageref{lastpage}} \pubyear{2013}

\maketitle

\label{firstpage}

\begin{abstract}

Depending on the topology of the magnetic field and characteristics of
turbulent motions, diffusion can significantly affect the 
distribution of elements, in particular helium, in the intracluster
medium (ICM). As has been noted previously, an incorrect assumption
about the helium abundance will lead to an error in the iron 
abundance determined from X-ray spectroscopy. The corresponding effect
on the temperature measurement is negligibly small. An incorrectly 
assumed helium abundance will also lead to a systematic error in
angular distance measurements based on X-ray and Sunyaev--Zeldovich
(SZ) observations of clusters of galaxies. Its magnitude is further
amplified by the associated error in the metal abundance determination,
the impact being larger at lower ICM temperatures. Overall, a factor
of $2$--$5$ error in the helium abundance will lead to an $\approx
10$--25\% error in the angular distance.

We solve the full set of Burgers equations for a multi-component
intracluster plasma to determine the maximal effect of diffusion on
the interpretation of X-ray and microwave observations of clusters of
galaxies. For an isothermal cluster, gravitational sedimentation can
lead to up to a factor of $\sim 5$--$10$ enhancements of helium and 
metal abundances in the cluster center on a $\sim3$--$7$~Gyr
timescale. In cool-core clusters on the contrary, thermal diffusion 
can counteract gravitational sedimentation and effectively remove
helium and metals from the cluster inner core.  In either case, a
significant, up to $\approx 40\%$, error in the metal abundances
determined by means of X-ray spectroscopy is possible. The angular 
distance determined from X-ray and SZ data can be underestimated by
up to $\approx 10$--$25\%$. 

\end{abstract}


\section{Introduction}
\label{secInt}

Clusters of galaxies are an important tool of observational
cosmology. The key role in fulfilling their potential as cosmological 
probes belongs to X-ray observations. X-ray imaging and spectroscopy, 
combined with the assumptions of hydrostatic equilibrium, symmetry
and uniformity, yield the total gravitating masses of clusters of
galaxies and their gas fractions. Another opportunity for cosmological
measurements with clusters of galaxies is provided by the
Sunyaev--Zeldovich (SZ) effect \citep{s1}. New-generation space- and 
ground-based SZ experiments, such as Planck, SPT, ACT and CARMA, are
capable of measuring electron pressure in the intracluster medium
(ICM) independently of X-ray data
\citep[e.g.][]{plag10,reese12,plag12,planck13}. Combined observations
in the microwave and X-ray bands allow one to measure the angular
distances to clusters of galaxies and thus to determine the Hubble
constant \citep{silk78,sunyaev80,carl02,m1,bon06,k1,planck12}. 
 
Interpretation of X-ray and SZ observations of clusters of galaxies is
subject to a number of uncertainties. The nature of many (but not all)
of these uncertainties are related to the physical state of the ICM,
for example to the assumptions of hydrostatic equilibrium or
symmetries in the ICM distribution, the role of non-thermal pressure, 
abundance distributions of elements etc. Many of these uncertainties
have been extensively discussed and are taken into account by
sophisticated data interpretation procedures
(e.g. \cite{bon06}). However, the effects of possible non-solar 
abundance distributions of elements remained so far largely ignored.

Diagnostics of element abundances in the ICM is based on measurement
of their line emission, which, given the characteristic ICM
temperatures, is in the X-ray band. This works reasonably well 
in principle for most of the cosmically abundant elements. However,
X-ray spectroscopy provides no information on the helium-to-hydrogen
ratio in the ICM, because both elements are fully ionized at these
temperatures and produce no spectral lines. Therefore in practice,
the helium abundance is usually assumed to be equal to its primordial
value. The latter is known quite accurately from the theory of Big
Bang nucleosynthesis which predicts the He fraction in the total baryonic
mass density of \mbox{$Y = 0.2482 \pm 0.0007$}
\citep{wal91,knel04}. However, abundances of elements in the ICM may
differ significantly from the primordial values. It has been shown
that sedimentation of helium  and heavy elements
\citep{fab1,g84,ch2,ch1,ettori06,sh1} may take place in the 
central regions of clusters. If transport processes are not
significantly suppressed in the ICM, helium abundance may increase by
a factor of 2 or more in the cluster center. For comparison, helium
enrichment by stars is not expected to be significant. 

An incorrect assumption about helium abundance in interpreting
observations of clusters of galaxies  will lead to several important
consequences. Firstly, the incorrect calculation  of the continuum
level, per particle, will result in an incorrect measurement of the
metal abundances and emission measures by means of X-ray spectroscopy 
\citep{drake98,ettori06}. That in turn will result in incorrect
estimates of the gravitating and gas masses of the cluster 
\citep{ettori06}. Secondly, incorrectly assumed helium abundance will
affect the results of angular distance measurements based on X-ray and  
microwave observations \citep{mar,bul}. 

Although different aspects of the helium abundance problem have 
already been discussed, the amplitude of its effects having been
estimated and their potential importance for cosmological measurements
with clusters of galaxies stressed out, none of the previous
treatments included the full consideration of the diffusion problem. In
particular, the effect of thermal diffusion was not taken into
account. The importance of the latter in considering cool-core
clusters has been demonstrated by \citet{sh1}. In particular, it was
shown that the temperature gradients in cool-core clusters are large
enough for thermal diffusion to counteract gravitational 
sedimentation and to reverse the flow of heavy particles, resulting in
the effective removal of helium and metals from the cluster inner
core.

The goal of the present paper is to consider the full set of Burgers
equations for a multi-component ICM plasma in order to estimate the
maximum possible impact of diffusion of elements in the ICM on the
interpretation of X-ray and microwave observations of clusters of 
galaxies. In the following we will consider both isothermal and
cool-core clusters. As a representative example of a cool-core cluster
we will use Abell 2029. The paper is organized as follows. In Section
2, we revisit the problem of bias in the observationally derived ICM
parameters caused by  incorrectly assumed helium abundance. In Section
3, we describe our treatment of the diffusion problem and present the
results of computations for A2029 and an isothermal cluster of the same
mass. Finally, in Section 4 we discuss implications of our results for
observational cosmology.

\section{Bias in X-ray and SZ derived quantities due to incorrectly assumed
helium abundance}
\label{secXSZ}

For a spherically symmetric gas distribution, the X-ray surface
brightness in the direction of the cluster center is given by:

\begin{equation}
S_x =  \frac{2}{4 \pi (1+z)^4} D_a \int\limits_0^{\theta_c}
n_e^2(\theta) \Lambda(A(\theta),T_e(\theta)) d\,\theta.
\label{eq1}
\end{equation}
Here, $z$ and $D_a$ are the cluster redshift and angular distance,
respectively, $n_e(\theta)$ and $T_e(\theta)$ are the electron number
density and temperature at distance $r = D_a \theta$ from the cluster
center along the line of sight, $\Lambda$ is the X-ray cooling
function of the ICM integrated over a given energy 
band, and $\theta_c$ is the cluster angular size. Note that we
define $\Lambda$ relative to $n_e^2$.  

\begin{figure}
\includegraphics[width=0.95\columnwidth]{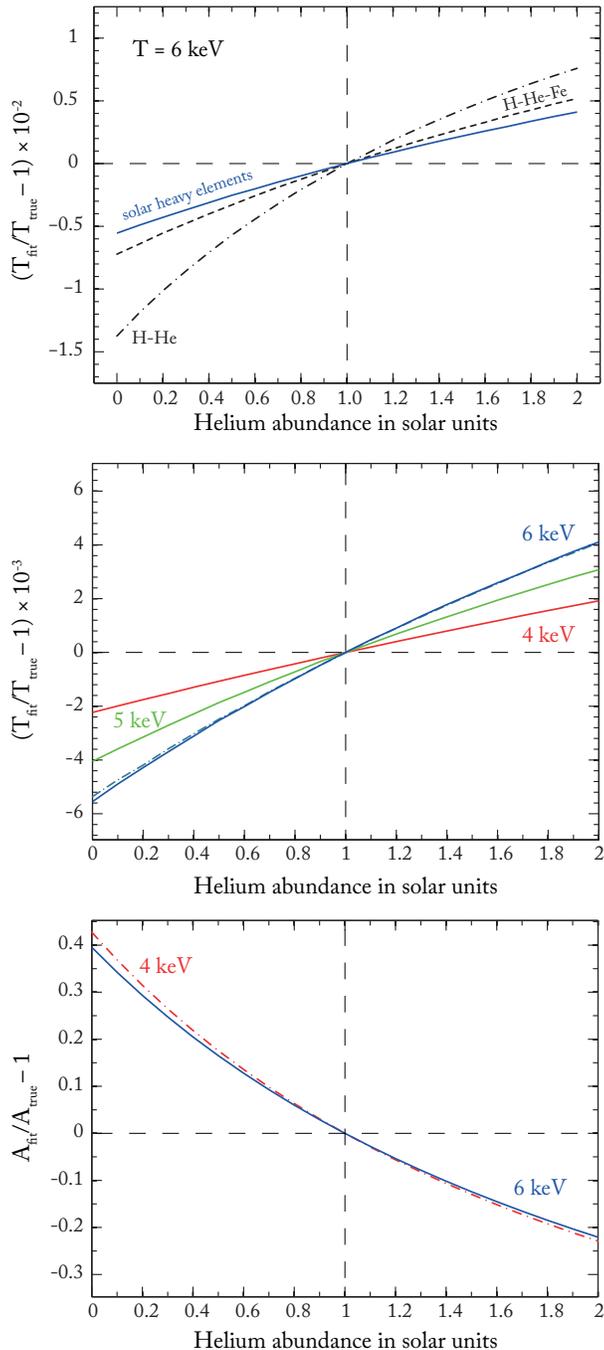}
\caption{Bias in the determination of the temperature (top and middle
  panels) and metal abundance (bottom) as a function of the helium
  abundance. The helium abundance was assumed to be solar in spectral
  fitting. {\it Top:} Temperature bias for gas with $kT=6$~keV and
  different compositions: solar abundances of elements heavier than
  helium (blue solid line); plasma consisting of H, He and
  solar-abundance Fe (black dashed line); H--He plasma (black
  dash-dotted line). {\it Middle:} Temperature bias for gas with solar
  abundances of elements heavier than helium and different
  temperatures: 4 (red), 5 (green) and 6 (blue) keV. The blue
  dash-dotted line corresponds to the bias for spectra generated with
  the XMM-Newton instrumental response ($T=6$~keV). {\it Bottom:} Bias
  in the metal abundance determination for gas with solar abundances
  of elements heavier than helium and temperatures of 4 (red) and 6
  (blue) keV.
} 
\label{figTA}
\end{figure}
% The  blue dash-dotted line corresponds to the bias for spectra generated with the XMM-Newton instrumental response, for $kT=6$~keV)
The amplitude of the nonrelativistic Sunyaev--Zeldovich effect is
proportional to the Comptonization parameter. The Comptonization
parameter for the same (central) line of sight as in
equation~(\ref{eq1}) is given by 
\citep{s1} 
\begin{equation}
y = 2 D_a \frac{k \sigma_T} {m_e c^2} \int\limits_0^{\theta_c}
T_e(\theta) n_e(\theta)  d\,\theta, 
\label{eq2}
\end{equation}
where $m_e$, $c$, $\sigma_T, k$ are the electron mass, speed of light,
Thomson cross section and Boltzmann’s constant, respectively. 

\subsection{X-ray derived quantities}
\label{secXray}

At typical ICM temperatures of $T \sim 10^7$--$10^8$~K, hydrogen and
helium are fully ionized and radiate mostly by bremsstrahlung. Hence,
the helium abundance cannot be determined directly from X-ray
observations. Analyzing the X-ray spectrum of the ICM, an observer
has to make some assumptions about the helium--hydrogen ratio in order
to  determine the abundances of heavy elements and the temperature and 
density of the gas. 

To estimate the bias in these quantities caused by an incorrect
assumption about helium abundance, we carried out the following
simulations. Using the VAPEC model \citep{sm} in XSPEC \citep{ar96},
we generated X-ray spectra for different gas temperatures, fixing the
abundances of heavy elements to their solar values and varying the
helium abundance from 0.1 to 2 in solar abundance units. In running the {\tt fakeit} command in XSPEC we do not use the counting statistics in order to exclude noise from the simulations. As the {\tt fakeit} command still assigns the formal statistical errors to the simulated spectrum, the subsequent fitting of the simulated spectrum is not compromised, except for chi-square value, which equals to zero.
The simulated spectrum was then fit by the same model with the helium abundance
fixed at the solar value and other element abundances characterized by
a single number $z$ --- the abundance ratio to the solar value, which,
along with the temperature, were free parameters of the fit. In the
simulations and spectral fitting, the Chandra (ACIS-I3 Cycle 15) 
response was used; the spectral fitting was performed in the 0.5--10
keV energy band. Poisson noise was not included in simulated spectra. 

The results of these calculations are presented in
Fig.~\ref{figTA}. In the top and middle panels, we compare the bias in
the determination of temperature for plasmas of different composition
and temperature. The bias is significantly stronger in the case of pure
H--He plasma compared to gas enriched by metals and increases with
temperature. However, its overall amplitude is negligibly small, less
than 1\%. Replacing the Chandra response function with that of
XMM-Newton EPIC MOS has practically no effect on this result. Hence,
the error in the helium abundance assumption does not significantly
affect the determination of gas temperature. This can be explained by
the fact that the inferred gas temperature is determined by the shape
of the spectral continuum and relative intensities of spectral
lines. Changing the helium abundance affects the continuum shape only 
slightly because the energy dependence of the Gaunt factor weakly
varies with the ion charge $Z$ (see, e.g., \citealt{hum88}), so that
hydrogen and helium have only slightly different shapes of
bremsstrahlung spectra.

In the bottom panel of Fig.~\ref{figTA}, we show the relative error in
the determination of heavy element abundances as a function of the
true helium abundance for different gas temperatures. The error in the 
determination of heavy element abundances resulting from the
assumption of solar helium abundance can reach $\sim 40 \%$ if the
true helium abundance is zero. For a factor of $2$ error in the
assumed helium abundance, the error on the heavy element abundances
is about $\approx 20\%$. Similar results were previously obtained by
\citet{drake98, ettori06}. The error in the heavy element abundances
arises because, due to incorrectly assumed helium abundance, the
spectral model incorrectly calculates the continuum level per hydrogen
atom, which affects the ratio of the emission line intensities to the
continuum. As increasing helium abundance leads to increasing the
continuum level per particle, the heavy element abundances are
underestimated when the helium abundance is underestimated. The effect
depends weakly on the gas temperature, being larger for lower
temperatures. 

%%%%%%%%%%%%%%%%%%%%%%%%
\begin{figure}
\includegraphics[width=\columnwidth]{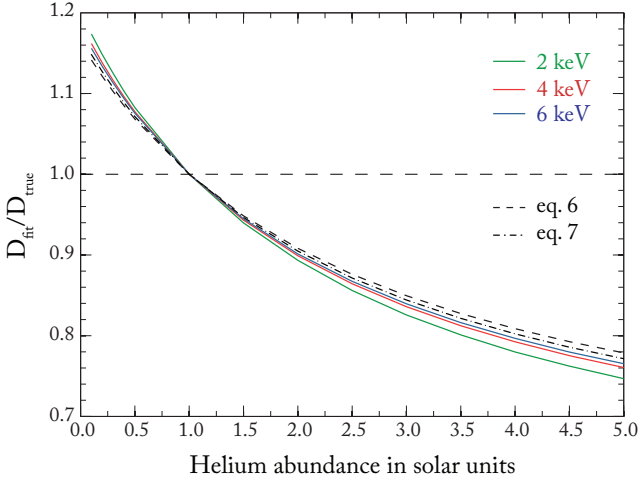}
  \caption{The bias in the distance to a uniform cloud of hot plasma
    derived from a combination of SZ and X-ray (Chandra) data, as a
    function of helium abundance. In the simulated data, the
    abundances of all elements heavier than helium were set equal to
    their solar values. The simulated data were interpreted assuming
    solar helium abundance, whereas the abundances of heavy elements
    were determined from X-ray spectral fitting. The green, red and
    blue solid lines show the distance bias computed using the precise
    expression, eq.~(\ref{eq7}), for the plasma temperature of 2, 4
    and 6 keV. For comparison, the black dashed and dash-doted lines
    show the bias for H--He plasma derived from the precise
    expression, eq.~(\ref{eq5}), and approximate relation,
    eq. (\ref{eq6}), respectively.} 
  \label{figDa}
\end{figure}

\subsection{Angular distance}
\label{secAng}

The X-ray brightness and SZ decrement depend on different powers 
of the electron density. Therefore, a combination of X-ray and 
SZ observations provides an opportunity to determine this density and 
convert it to the cluster distance using equations~(\ref{eq1}) and
(\ref{eq2}) \citep[]{sunyaev70,silk78,sunyaev80}:  
\begin{equation}
D_a =  \frac{y^2}{S_x} \frac{\int\limits_0^{\theta_c} n_e^2\, \Lambda
  d \theta}{\left( \int\limits_0^{\theta_c} n_e\, T_e d\theta
  \right)^2} \frac{m_e^2 c^4}{2 k^2 \sigma_T^2 4 \pi (1+z)^4},
\label{eq3} 
\end{equation}
where $y$ and $S_x$ are the Comptonization parameter (determined from
the observed SZ decrement) and X-ray surface brightness, both
measured in the direction towards the cluster center. Whereas both $y$ 
and $S_X$ are directly measured quantities, the radial temperature
profile $T(\theta)$ and the shape of the density profile $n_e(\theta)$
can  be inferred from observations, e.g., from spatially resolved
broadband spectroscopy. Note that in order to determine the
normalization of $n_e(\theta)$ one would need to know the distance to
the cluster, however, the density normalization is not required for
the angular distance calculation as it cancels out in
eq.\ref{eq3}. The cooling function $\Lambda$ depends, apart from the
temperature $T$, on the assumed helium abundance $x$ and metal
abundances $z$, the latter usually being  determined from X-ray
spectral fitting, i.e. $\Lambda=\Lambda(x,z,T)$. Note that the cooling 
function is normalized to the square of electron density $n_e^2$. 

To investigate how distance measurements can be affected by
an error in the assumed helium abundance, we considered an isothermal
spherical gas cloud of constant density and homogeneous chemical
composition with an angular size of $\theta_c$ located at distance
$D^{true}_a$ from the observer. In this case, eq.(\ref{eq3}) can be
reduced to:  
\begin{equation}
\frac{D^{fit}_a}{D_a^{true}} = \left(\frac{T^{true}}{T^{fit}}\right)^2
\frac{\Lambda(x^{as},z^{fit},T^{fit})}{\Lambda(x^{true},z^{true},T^{true})} \approx
\frac{\Lambda^{as}}{\Lambda^{true}}, 
\label{eq7} 
\end{equation}
where $T^{true}$ and $x^{true}$, $z^{true}$ are the true temperature  and
abundances of the gas, $T^{fit}$ and $z^{fit}$ are the corresponding
best-fit values obtained from X-ray spectral fitting and $x^{as}$ is
the assumed helium abundance and $D^{fit}_a$ is the angular distance
inferred from the analysis. 

For hydrogen--helium plasma ($z=0$), the cooling function is given by:  
\begin{equation}
 \Lambda (x,z=0,T)=\epsilon_{ep}(T)\ 
\frac{1 + 4x\,g_{ff}(Z_{He})/g_{ff}(Z_H)}{1 + 2x}, 
\label{eq4}
\end{equation}
where $\epsilon_{ep}(T)$ is the bremsstrahlung emissivity for a pure
electron-proton plasma, $x = n_{He}/n_H$ is the helium abundance by
the number of particles with respect to hydrogen and $g_{ff}(Z)$ is
the frequency-averaged Gaunt-factor for the element with charge
$Z$. The relative error in the distance determined from
equation~(\ref{eq7}) is then 
\begin{equation}
\frac{D^{fit}_a}{D_a^{true}} =
\left(\frac{T^{true}}{T^{fit}}\right)^2 \frac{1 + 4 \tilde{g}\, x^{as}}{1 + 4 \tilde{g}\, x^{true}} \frac{1 + 2 x^{true}}{1 + 2
  x^{as}}.
\label{eq5}
\end{equation}
where  $\tilde{g} \equiv g_{ff}(Z_{He})/g_{ff}(Z_H)$. If we neglect
the weak dependence of the Gaunt-factor on $Z$ \citep{hum88} and the
related weak temperature bias (see Fig.~\ref{figTA}), we can simplify
the above expression as follows:  
\begin{equation}
\frac{D^{fit}_a}{D_a^{true}} \approx \frac{1 + 4 x^{as}}{1 + 4 x^{true}}
\frac{1 + 2 x^{true}}{1 + 2 x^{as}},  
\label{eq6}
\end{equation} 
This approximation was previously derived by \cite{mar}. The distance
bias described by equations~(\ref{eq5}) and (\ref{eq6}) is shown in
Fig.~\ref{figDa}. We see that the difference between
equations~(\ref{eq5}) and (\ref{eq6}) is small.

If the gas contains heavy elements, the dependence of the distance bias
on $x^{as}$ becomes more complicated. The heavy elements contribute to the
total X-ray luminosity by continuum emission (recombination,
bremsstrahlung, 2-photon emission) and line emission. As was shown in
Section~\ref{secXray}, for $x^{as}\neq x^{true}$, bias also appears in
measuring the heavy element abundances. Hence, there is an additional
bias in $D_a^{fit}$ due to the incorrectly accounted contribution of
heavy elements to $\Lambda$ in equation~(\ref{eq3}). 

To investigate this quantitatively, we used the results of
simulations performed in Section~\ref{secXray}. From the best-fit
values of temperature and metal abundances obtained there, we computed
the cooling function $\Lambda^{as}$ in eq.(\ref{eq7}) and the ratio
$D_a^{fit}/D_a^{true}$.  The result is shown in Fig.~\ref{figDa}.  
Replacing the Chandra response with that of XMM-Newton yields
practically the same results. 

As can be seen from Fig.~\ref{figDa}, for helium--hydrogen plasma
there is no significant difference between the accurate
eq.(\ref{eq5}) and approximate eq. (\ref{eq6}). However, as is clear from comparison of dashed to solid lines in  Fig.~\ref{figDa}, formula~(\ref{eq6}) underestimates the effect for plasma of solar abundance by $\approx 5$--10\% at large value of helium abundance, the difference becoming larger at lower temperatures. 

\section{Redistribution of helium and heavy elements by diffusion in
  clusters of galaxies}  
\label{secDif}

The element abundance distribution in the ICM is determined by a
number of physical processes. The initial composition of the ICM will
change with time due to ejection of heavy elements by supernovae,
turbulent mixing and diffusion in the gas. The role of diffusion in
shaping ICM abundance profiles has been disputed for a long
time. As is well known, transport processes in the ICM may be
suppressed by magnetic fields
(e.g. \citealt{g84,ettori00,mar03,kom13}). However, the
chaotic fluctuations of the magnetic field produced by turbulence can
make the large scale transport coefficients big enough to make the
diffusion important \citep{nar}. On the other hand, turbulent  
mixing can counteract diffusion, its effect on the distribution of
elements depending on the turbulence spatial scale and velocity
\citep{asc}.

We considered the problem of diffusion of elements in the ICM 
in full detail but ignored the complexity introduced by the magnetic 
fields, turbulent mixing and enrichment of the ICM by supernovae.
The goal of this exercise was to estimate the maximal effect that
diffusion can have on various cosmologically important measurements 
with clusters of galaxies.  

\subsection{Calculation method}
\label{secCalc}

Diffusion of elements is driven by the force of gravity and by density
and temperature gradients. Density gradients tend to restore a
homogeneous distribution of the element abundances in plasma, whereas
gravitational sedimentation tends to concentrate plasma's heavy
particles to the cluster center; the equilibrium state distribution of
elements of mass $m_i$ is proportional to the Boltzmann factor $n \sim
e^{-m_i \phi(r)/kT}$, where $\phi$ is the gravitational potential
\citep{g84,ch1,abram}. The presence of temperature gradients in plasma
gives rise to thermal diffusion, which tends to remove heavy and more
highly charged particles from colder regions
\citep{bur1,ch70,mon85,sh1}.

\begin{figure}
\includegraphics[width=\columnwidth]{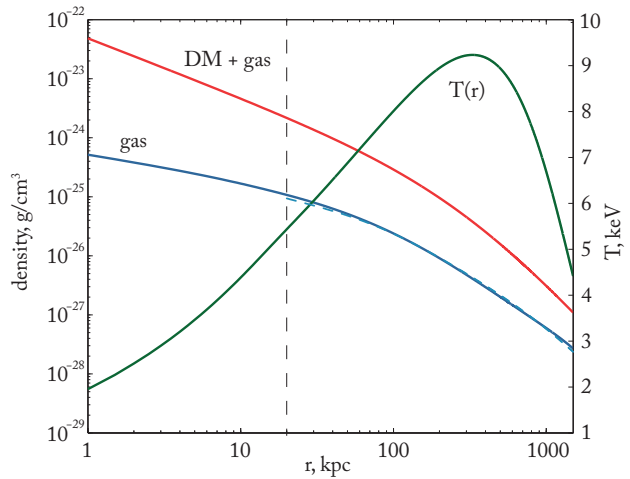}
\caption{The profiles of the ICM density (blue), temperature (green) and
  total mass density (red) in Abell 2029 that were used in the
  diffusion calculations. For the temperature profile, we combined the 
  models of \citep{vih2} and \citep{lew} outside and inside 20~kpc
  (vertical dotted line), respectively. The radius of 20~kpc
  corresponds to about $14^{\prime\prime}$.  The blue dashed line shows the ICM density profile from \citealt{vih2}.}
 \label{figA2029}
\end{figure}

\begin{figure*}
  \includegraphics[width=\textwidth]{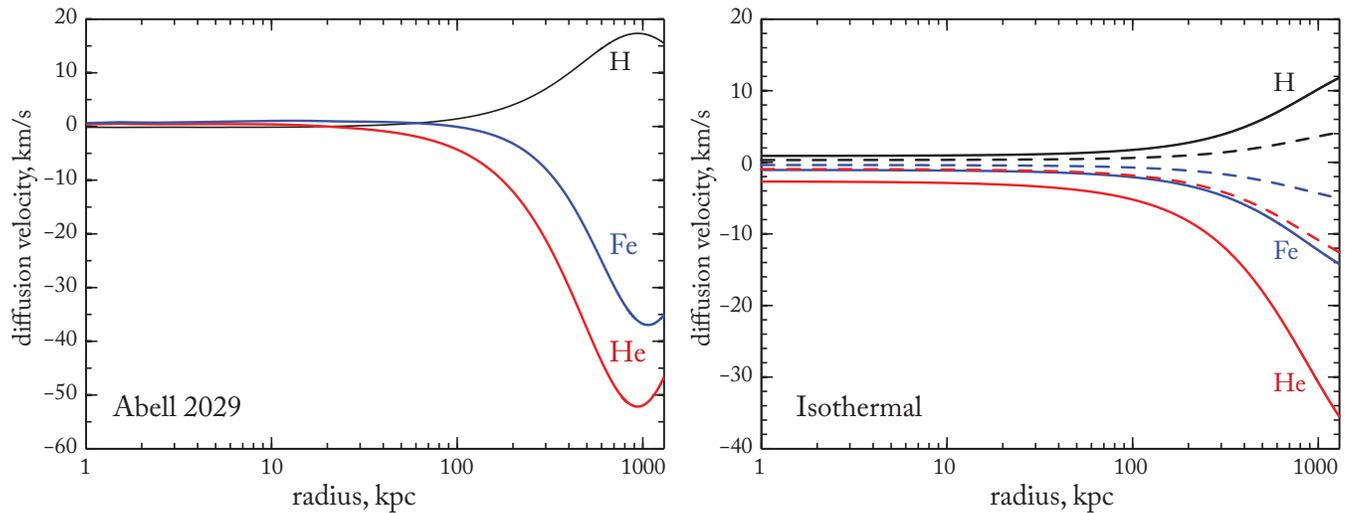}
  \caption{The initial diffusion velocity of H$^+$ (black), He$^{+2}$
    (red) and Fe$^{+24}$ (blue). Positive velocity corresponds to
    outward flow. The left panel shows the results for A2029 and the
    right panel for an isothermal cluster with a temperature of 3
    (dashed line) and 6 (solid line) keV. The mass profiles
    of the isothermal clusters are the same as for A2029.}
  \label{figDVel}
\end{figure*}
\begin{figure*}
\includegraphics[width=\textwidth]{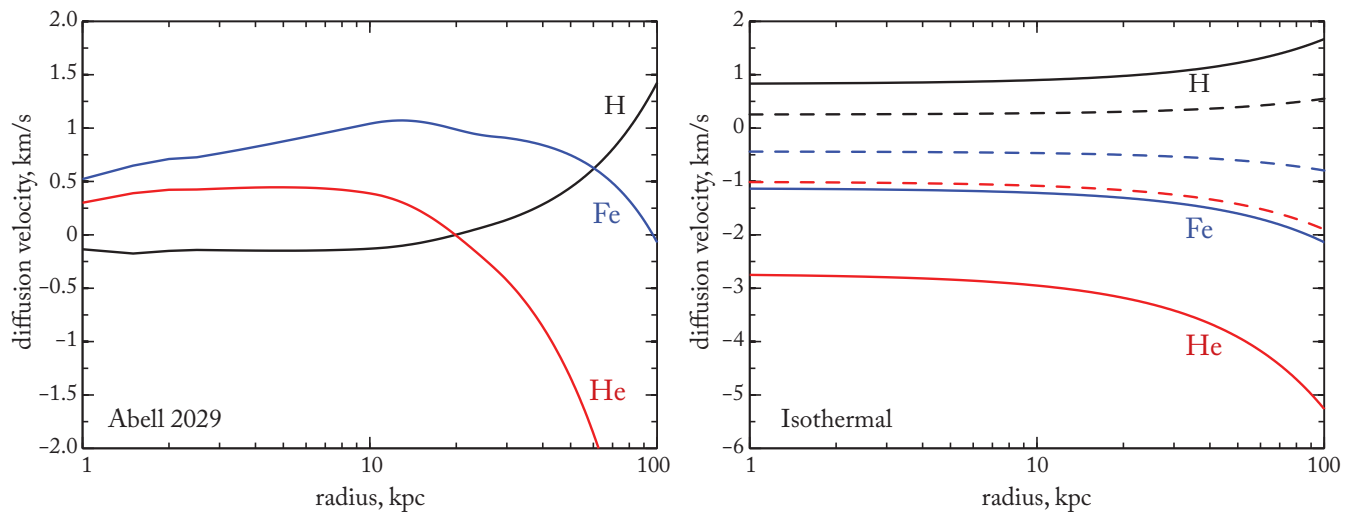}
  \caption{Same as Fig.~\ref{figDVel}, but for the central parts of
    the clusters.} 
  \label{figDVel_c}
\end{figure*}

To take into account all these processes, we considered a full
diffusion problem by solving the system of Burger’s equations
\citep{bur1}. We used the numerical scheme from \cite{sh1}; the 
interested reader is referred to that paper for the details. Here we
describe the main simplifying assumptions: 

\begin{enumerate}

\item We consider a 4-component plasma consisting of hydrogen,
  helium, a heavy element ($\mathcal{A}$, $Z$) and electrons. The 
  hydrogen and helium are assumed to be fully ionized. In all
  of the calculations presented here, the heavy element is
  Fe$^{+24}$. As was demonstrated in \citet{sh1}, adding other
  heavy ions does not lead to noticeable differences in the diffusion
  picture, moreover, distributions of all metals change in an
  approximately identical manner. 

\item We use a spherical model of a cluster in hydrostatic
  equilibrium. Because of the non-negligible helium abundance,
  diffusion changes the total pressure: $p = \Sigma n k_b T$, hence
  hydrostatic equilibrium is violated and a net flow of particles 
  appears. However, because the sound speed in the ICM is much higher
  than the diffusion velocity, hydrostatic equilibrium restores
  quickly and stationary Burger's equations remain valid. The (small)
  net flow velocity is calculated using the Euler equation.

\item All velocities, temperature and density gradients are required
  to vanish at the inner point $r=0$. At the outer boundary, a constant
  density boundary condition is set. This is equivalent to assuming
  that the galaxy cluster is imbedded in an infinite reservoir of
  gas. In our baseline model, the cluster outer boundary is located at
  a distance of $1500$ kpc from the cluster center. We have verified
  that the solution in the inner regions of the cluster is insensitive
  to this condition (see also \citealt{sh1}). Furthermore, we also
  considered the case of an opaque boundary condition by setting the
  speeds of all species equal to zero at the outer boundary. The
  resulting solution at $t = 7$ Gyr differs by less than a few per
  cent from the solution for our default boundary conditions almost
  everywhere out to a radial distance of $\sim 1000$ kpc.
 
\item The evolution of the temperature profile is not
  self-consistently included in the calculations. The temperature
  profile is fixed, $T(r,t) = T(r,0)$, and heat transport in the 
  plasma is switched off, i.e. there is no heat flux term in our
  equations. This is equivalent to assuming that heat transport
  and the activity of the central AGN are fully compensated by the
  radiative cooling of gas. The possibility of a stable configuration
  of this type has been demonstrated e.g. in \cite{guo08}.
\end{enumerate}

\begin{figure*}
  \includegraphics[width=\textwidth]{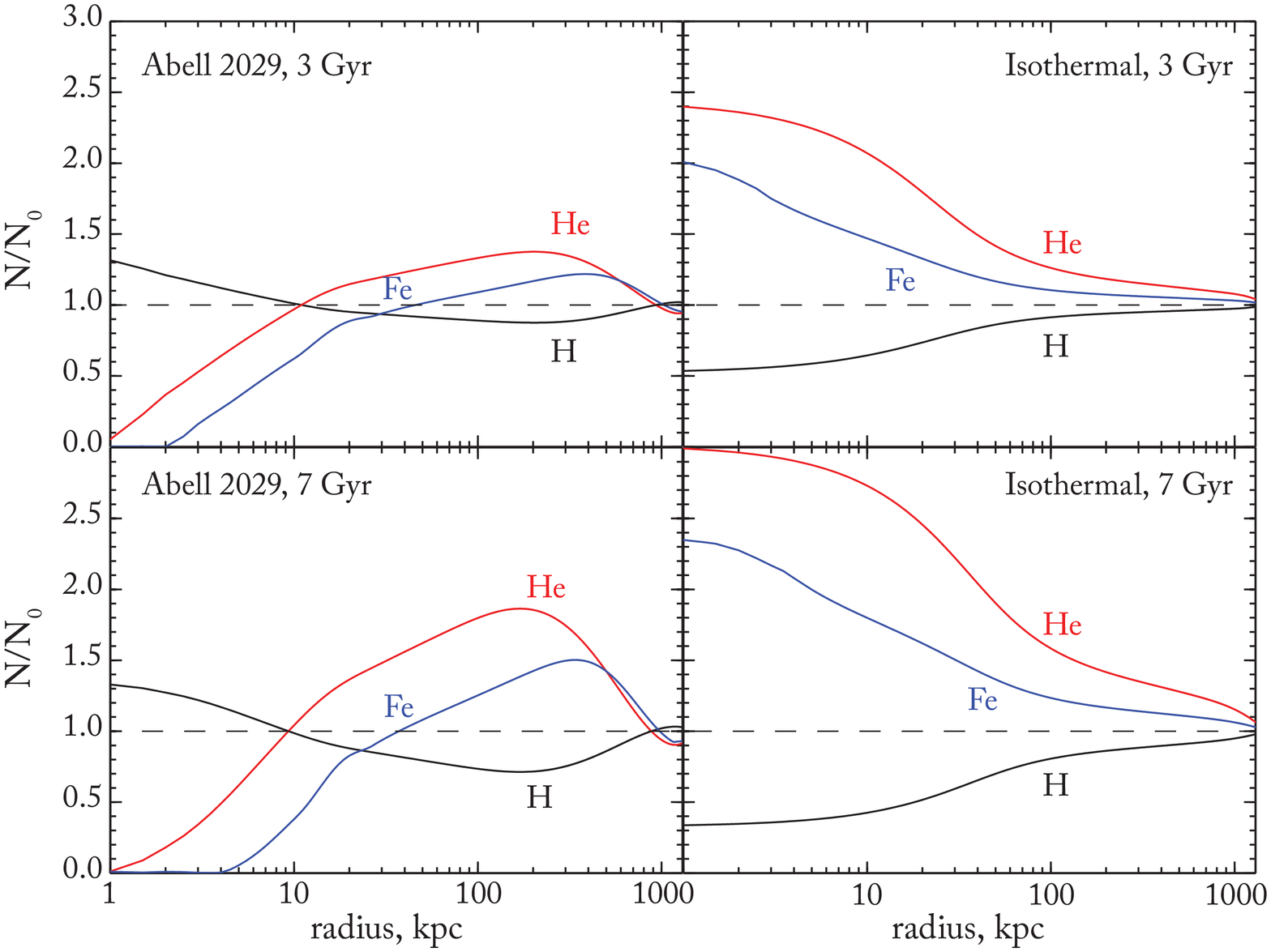}
  \caption{Element distributions after 3 (top panels) and 7 (bottom
    panels) Gyr of diffusion, normalized to the respective
    distributions at $t=0$. Hydrogen, helium and iron are shown the
    black, red and blue lines, respectively. The left panels are for  
    A2029 and the right panels are for an isothermal cluster with $kT 
    = 6$ keV and the same dark matter distribution as in A2029.} 
  \label{figAbund}
\end{figure*}

\subsection{Cluster models}

We conducted our calculations for cool-core and isothermal cluster
models. To build the cool-core cluster model, we use  data for the A2029
cluster of galaxies. This cluster, located at $z = 0.0767$, is known
to have regular X-ray morphology and has been studied extensively in
X-rays with ROSAT \citep{sar1}, Chandra \citep{lew,vih1,vih2} and
XMM-Newton \citep{sn}.

We use the best-fit model from \cite{vih2} to describe the temperature 
profile of A2029. This model is defined outside of 20~kpc from 
the cluster center. To continue the temperature profile inward of 20~kpc, 
we use the best-fit model from \cite{lew}, which is defined down to
1~kpc from the cluster center. We smoothly stitched the two models at
20~kpc using a 5-degree polynomial. Within 1 kpc of the cluster
center, we extrapolated the temperature profile using a 3-degree
polynomial with a vanishing derivative in the cluster center. The
central temperature is fixed at 2~keV. 

For the total mass, we use a NFW \citep{nfw97} profile 
\begin{equation}
\rho = \frac{\rho_s}{r/r_s (1 + r/r_s)^2},
\label{eq8}
\end{equation} 
with $r_s = 337$ kpc \citep{vih2}. The normalization value, $\rho_s
=1.7 \times10^{-25}\ \mbox{g/cm}^3$, was determined using the best-fit
gas density profile from \cite{vih2}, the aforementioned temperature
profile and the condition of hydrostatic equilibrium, $\frac{\Delta
  p}{\rho} = -g$. Using the temperature profile, the normalized NFW
profile and the condition of hydrostatic equilibrium, we finally
determined the gas density distribution. The resulting profile is
close to that from \cite{vih2}. The gas mass fraction is 15\% inside
$r_{500} = 1360$ kpc (recall that the outer boundary is fixed at 1500
kpc). The final profiles of the total mass density, gas density and
temperature are shown in Fig.~\ref{figA2029}.

For the isothermal cluster models, we assume the same dark matter
distribution as in A2029.  We considered two isothermal models 
with temperatures of 3 and 6~keV. Similar to the cool-core model, the
gas density distribution was calculated assuming hydrostatic
equilibrium. The gas mass fraction in these models is also 15\%. The
isothermal cluster models allow us to isolate the effect of thermal
diffusion, effectively switching it off in Burger's equations.

In all the considered cluster models, the mean free path for protons
at the outer radius is $\sim 10$--20~kpc, hence the diffusion
approximation is valid over the entire cluster.

\subsection{Evolution of element abundance profiles}
\label{secNum}

We start our calculations with the solar mass fractions of $0.75$,
$0.25$, $1.8 \cdot 10^{-3}$ for H, He and Fe, respectively
\citep{anders}. The initial abundance profiles are assumed to be
flat. The initial profiles of the diffusion velocity of H$^{+}$,
He$^{+2}$ and Fe$^{+24}$ for A2029 and the isothermal cluster are
shown in Fig.~\ref{figDVel}; Fig.~\ref{figDVel_c} zooms in on the 
velocities in the central part of the cluster. Positive velocity
corresponds to outflow and negative to inflow. 

In the central 20~kpc region of A2029, helium and iron have positive
velocities and only hydrogen flows towards the cluster center. Such
behavior is caused by the large temperature gradient in the central
part of the cluster (from the center to $\sim 300$~kpc,
Fig.~\ref{figA2029}). Thermal diffusion tends to evacuate heavy
elements from the cool core, counteracting gravity. On the other hand, the
outward decrease of temperature beyond $r > 500$~kpc acts in the same
direction as gravitational sedimentation and  increases the  diffusion 
velocity as compared to isothermal cluster models. 

Fig.~\ref{figAbund} shows the accumulated effect of the diffusion
after 3 and 7 Gyr. After 7 Gyrs, the thermal diffusion completely
removes iron from the central part of A2029. The same is true for
helium, albeit to somewhat lesser extent. The combined effect of
gravitational sedimentation and thermal diffusion increases the
concentration of both elements  at $\sim 100$--$200$ kpc from the 
cluster center. The hydrogen distribution changes in the opposite
sense, in order to keep hydrostatic equilibrium. As a result, complex
abundance distributions of elements are formed. The particular shape
of these distributions is determined by the detailed shape of the
temperature profile, as the net diffusion velocity depends on the
amplitude of the temperature gradients. Although the temperature
profiles vary significantly from cluster to cluster, a picture
qualitatively similar to that for A2029 arises  for other cool-core
clusters: the abundance of elements heavier than hydrogen is
suppressed in the cluster center and peaks at intermediate radii,
$\sim$ few hundred kpc, due to the combined effect of thermal
diffusion and gravitational separation (see \cite{sh1} for a few other 
examples). 

For the isothermal cluster model, the gravitational sedimentation
modifies the element distributions in a more monotonic fashion,
increasing their concentrations towards the cluster center. As
demonstrated in previous work \citep{g84,ch1}, the enhancement  is
most significant for helium and is somewhat smaller for iron and other
metals. In particular, the helium concentration in the cluster center
can increase by up to a factor of $\approx 2.4$--$3$ after $3$--$7$
Gyrs. Because of the factor of $\sim 2$--$3$ decrease of the hydrogen
concentration in the center, the increase in the helium abundance is
larger and can achieve a factor of $\sim 5$--$9$. As the diffusion 
velocity depends on temperature as $v\propto T^{3/2}$, the effect is
stronger for higher temperature clusters.

\subsection{Outflow and inflow of particles through the cluster outer
  boundary} 
\label{secOutflow}

Both theory and cosmological simulations predict that the temperature
profiles of clusters should become similar when radii are scaled to the cluster virial radius (for a detailed
discussion see, e.g., \citealt{kat93,lok02,bry98}). This prediction is
confirmed by observations \citep[e.g.][]{vih1}. A prominent feature
of the predicted and observed temperature profiles is that the ICM
temperature decreases outwards in the outer regions of clusters,
beyond several hundred kpc. As was demonstrated in the previous
sections, the combined effect of gravitation sedimentation 
and thermal diffusion will lead to inflow of elements heavier than 
hydrogen and outflow of hydrogen through the outer boundary of the
cluster of galaxies. This will result in a long-term increase of the
cluster-averaged abundances of elements heavier than hydrogen. 
Below,
we calculate the net flux of particles through the outer boundary
in order to evaluate the amplitude of this effect. Recall that in our
standard A2029 model, $r_{out} = 1.5$~Mpc. 

\begin{figure}
  \includegraphics[width=\columnwidth]{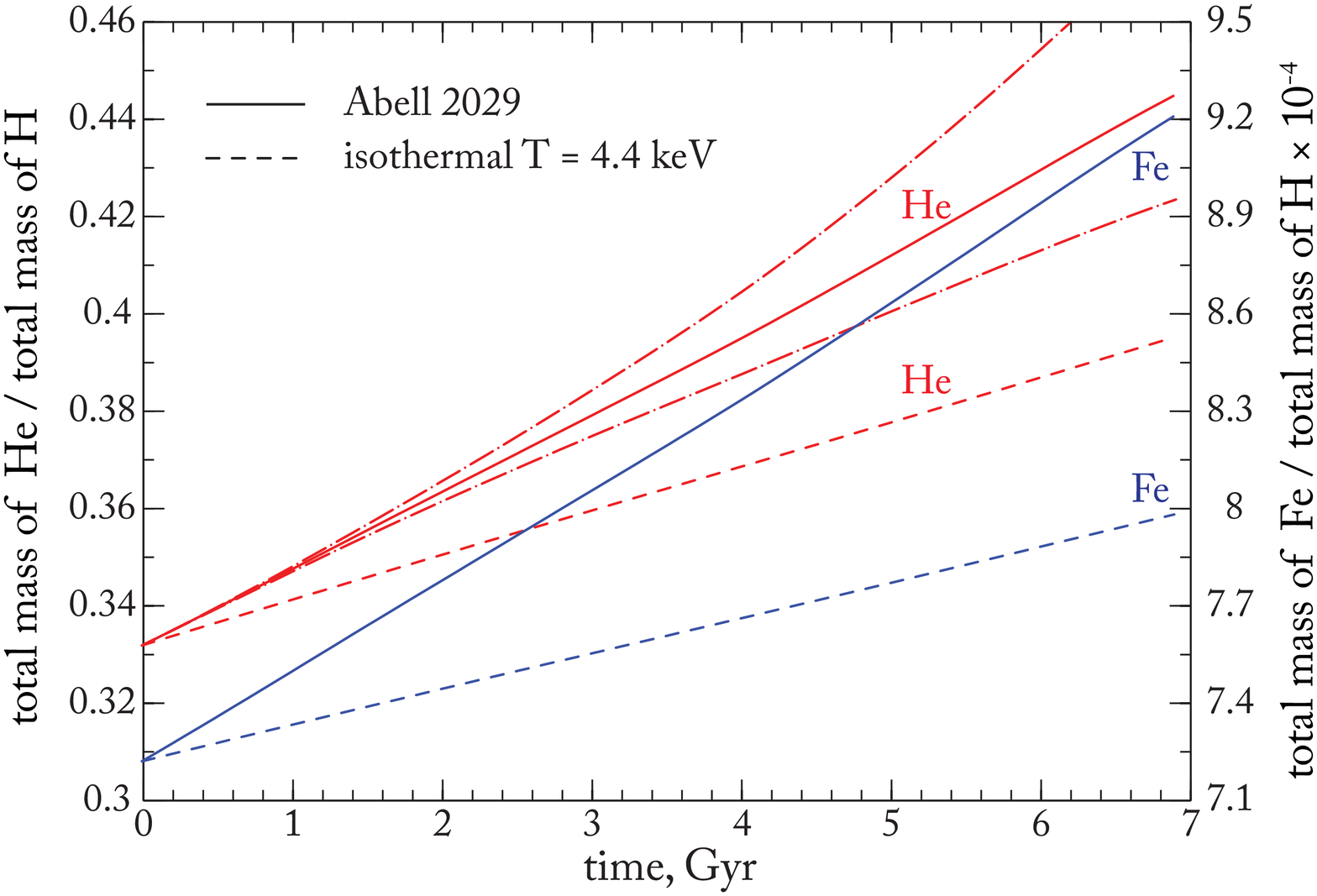}
  \caption{Evolution of the cluster averaged abundances, by mass, of
    helium (red lines) and iron (blue lines) as a function of time for
    A2029 (solid lines) and isothermal cluster with the temperature
    equal to the outer temperature of the A2029 model ($T=4.4$ keV)
    (dashed lines). The dash-dotted lines illustrate, how the bulk motion of gas on the outer boundary of the cluster, with velocity of $+40$ (outflow, upper curve)  and $-40$ km/s (inflow, lower curve), affects the evolution of helium abundance.}
  \label{fig:abund}
\end{figure}

In our cool-core cluster model, the ICM mass inside $r_{out}$ obtained
by integrating $\rho_g$ (Fig.~\ref{figA2029}) is $1.4\times
10^{14}$~\msun. At $t = 0$, the hydrogen, helium and iron mass
fractions take their solar values of $0.75$, $0.25$ and
$1.8\times10^{-3}$, respectively, and the corresponding diffusion
velocities at the outer boundary are $v_H = +13.8$, $v_{He} = -41.2$ 
and $v_{Fe} =-31.8$ km~s$^{-1}$ (Fig.~\ref{figDVel}). Hence,  the
initial mass flow rates are $q_H = -1190$, $q_{He} = 1185$ and $q_{Fe} = 5$
\msun\,/year, where positive sign means that the total mass of the
given element in the cluster is increasing. Thus, at the time $t=0$
the abundances will be increasing at rates of: 
\begin{eqnarray}
\dot{A}_{He}/A_{He}\approx 4.8\cdot 10^{-2} {\rm ~Gyr}^{-1}\\
\label{eq:adot_he}
\dot{A}_{Fe}/A_{Fe}\approx 3.9\cdot 10^{-2} {\rm ~Gyr}^{-1}
\label{eq:adot_fe}
\end{eqnarray}
In the isothermal cluster model, these rates are correspondingly
smaller. 

The detailed long-term evolution of the cluster-averaged abundances
computed from results of our numerical calculations is shown in
Fig.~\ref{fig:abund}. After 7~Gyr, 4.5\% of the hydrogen initially
present in the cluster will have outflow through the outer boundary,
whereas the total masses of helium and iron in the ICM will have
increased by 28\% and 22\%, respectively. This will result in an
increase of their cluster-averaged abundances by 33\% and 27\%,
respectively. For the isothermal cluster ($T=4.4$ keV), the
cluster-averaged abundances of helium and iron will have increased
by 19 and 11\%, respectively, after 7 Gyr. 
Due to outflow of hydrogen and inflow of helium and heavier elements through the outer boundary
of the cluster, the mean molecular weight of the ICM
increases with time. We note that due to our assumption of
hydrostatic equilibrium at $t = 0$, there is no net mass
flow through the outer boundary at the initial moment. Once
diffusion starts to operate, the total pressure changes and
the gas is pushed out of the equilibrium which is restored
via net gas motion. As a result, there is net inflow of mass
through the outer boundary and the total mass of the ICM
increases at the average rate of   $\sim 0.5\%$ per Gyr.
  
We conclude that the cluster-averaged abundances of helium and
metals can significantly increase due to the combined effect of 
thermal diffusion and gravitational sedimentation at the outer
boundary of the cluster, provided that these elements are 
present in sufficient amounts in the intergalactic medium (IGM),
outside this boundary. In the above calculations we assumed that
their abundances are solar in the IGM. This is true for
helium, but metals are not expected to be present in significant
amounts outside clusters of galaxies. Therefore, eq.(\ref{eq:adot_fe})
may strongly overestimate the effect for iron. 

\begin{figure*}
  \includegraphics[width=\textwidth]{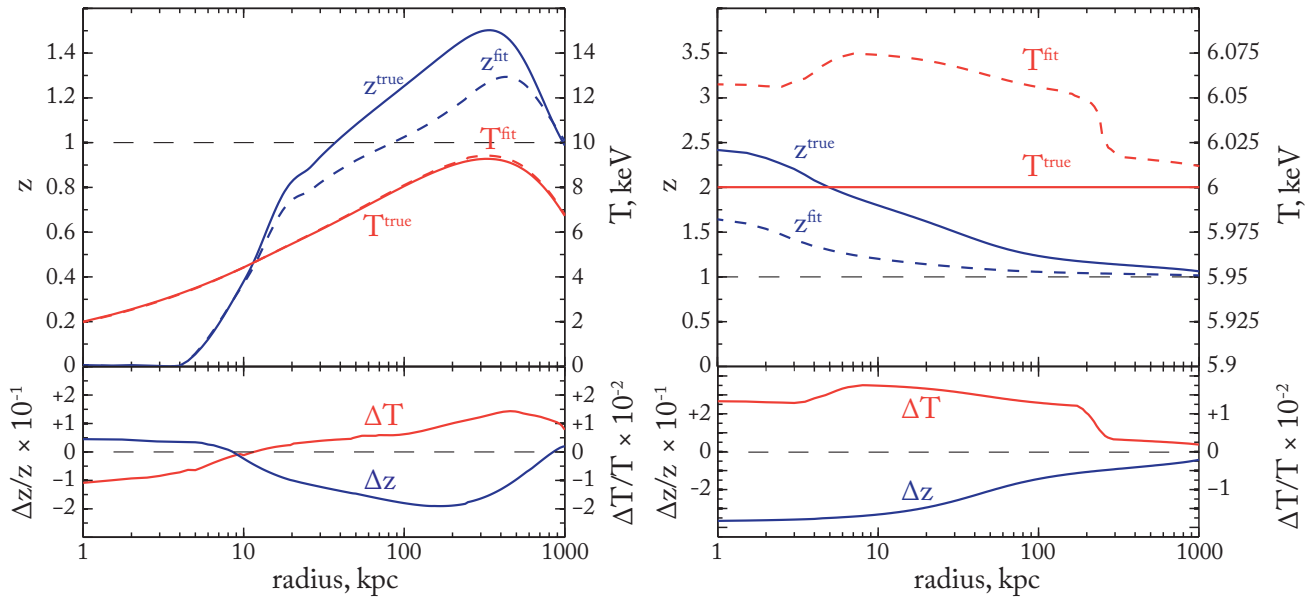}
  \caption{Abundance and temperature biases in X-ray spectral analysis
    for A2029 ({\em left}) and for an isothermal  cluster  with
    $kT=6$ keV ({\em right}) at $t=7$ Gyr
    (cf. Fig.~\ref{figAbund}). {\it Top:} The blue solid line shows
    the true metal abundance, while the blue dashed line shows its
    best-fit value obtained under the assumption of solar helium
    abundance. The corresponding red lines show the true and best-fit
    temperature profiles. {\it Bottom:} The relative error in the metal
    abundance (blue) and temperature (red) determination.}
\label{figTFe}
\end{figure*}

In these calculations, we did not account for bulk motion of gas at the outer boundary of the cluster. The full consideration of the effect of such motions is beyond the scope of this paper. Nevertheless, it can be easily shown that in the case of the gas outflow with the moderate velocity, $\la 100$ km/s, the amplitude of the discussed effect  will increase, whereas in the case of the gas inflow, the effect will be somewhat decreased. This is illustrated by the dash-dotted curves in Fig.~\ref{fig:abund}, showing the case of the gas bulk motion with velocity of $\pm 40$ km/s. Note that at significantly higher velocities the total mass of gas out(in) flow from the cluster becomes comparable to the total initial mass of the gas in the cluster, therefore full self-consistent consideration of the problem becomes essential.

We finally note that
the topology of the magnetic field and characteristics of turbulence
may be different in the central parts of clusters of galaxies and in 
their outskirts. Therefore, diffusion may be suppressed in the center
of the cluster but operate at sufficient strength in its outer regions
and vice versa.

\section{Effect of diffusion on the interpretation of X-ray and SZ
  data}

Thermal diffusion can thus significantly affect the distribution of
helium and heavy elements in galaxy clusters. This, together with the
bias in the determination of heavy element abundances introduced by
the standard assumption of solar helium abundance
(Section~\ref{secXSZ}), can have important consequences for the
interpretation of X-ray and SZ observations of clusters. We now use 
results of previous sections to estimate the expected biases
in the determination of cluster parameters.

We divided the model cluster into concentric spherical shells and
generated the X-ray spectrum of emission from each shell using the
APEC model in XSPEC and the electron number density, temperature and
element abundances resulting from the diffusion calculations in
Section~\ref{secNum}. We then fitted the generated spectra by APEC 
assuming the solar abundance of helium. The results are shown 
in Fig.~\ref{figTFe}. In the top panel, we compare the true radial
profiles of metal abundance and temperature with those determined from
the fits. Their fractional  difference is shown in the bottom panel; it
illustrates the bias in the determination of ICM parameters from X-ray
spectroscopy. These results are presented for spherical shells, rather
than for projected quantities, i.e. we implicitly assume that the
X-ray data are of sufficient quality to enable the de-projection
analysis. 

We see that the assumption of solar helium abundance results in  a
significant, up to $\approx 20$--$35\%$ bias in the metal abundance
determination. For the isothermal cluster it leads to a significant,
more than a factor of $\sim 2$, underestimate of the effect of the
gravitational sedimentation for metals. The total mass of heavy
elements in the ICM is underestimated by  5.7\% and 5.5\% for A2029
and the isothermal cluster, respectively. The 
total mass of gas is underestimated by 7\% and 8\%, respectively. On
the other hand, the temperature is only weakly sensitive to the helium
abundance assumption and bias in its determination is thus small.

\begin{figure*}
  \includegraphics[width=\textwidth]{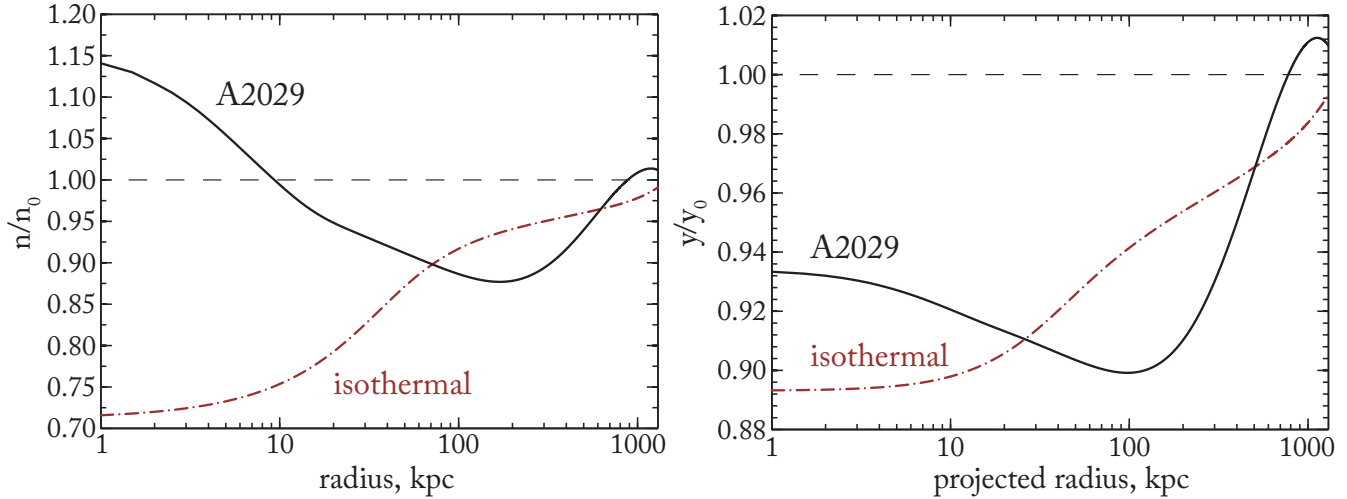}
  \caption{{\it Left:} The change in the radial profile of electron
    density after 7~Gyr of diffusion for A2029 (black line) and for
    the isothermal cluster with $kT=6$~keV (red line) as a function of
    the radial distance to the cluster center. {\it Right:} The
    corresponding change in the Comptonization parameter,  i.e. of the 
    amplitude of the SZ-effect, as a function of projected radius.} 
  \label{figSZ}
\end{figure*}

The redistribution of elements in the ICM also modifies the electron
density profile (due to the electroneutrlity condition) and 
consequently the amplitude of the SZ effect. In 
the left panel of Fig.~\ref{figSZ}, we show the change in the radial
profile of electron density after 7~Gyr of diffusion for the isothermal
cluster with $kT = 6$ keV and for A2029. Comparison with
Figs.~\ref{figAbund} and \ref{figSZ} indicates that the change in the
electron density is primarily caused by the diffusion of hydrogen. In
the right panel of Fig.~\ref{figSZ}, we show the resulting changes in
the Comptonization parameter as a function of the projected radius. 

\begin{figure*}
  \includegraphics[width=\textwidth]{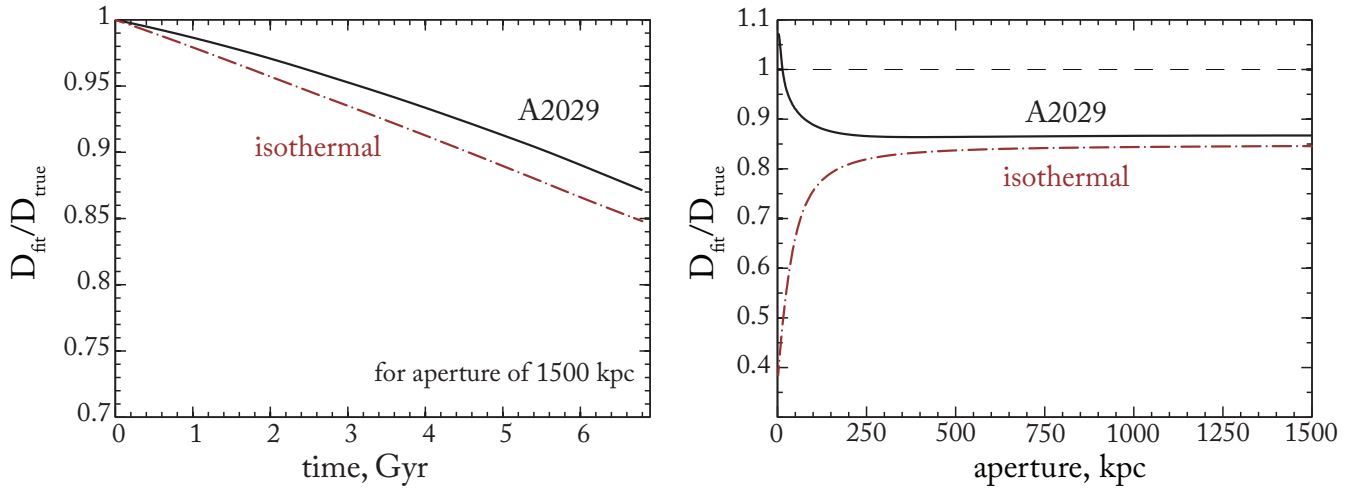}
  \caption{The bias in the angular distance  determined  from the
    combined analysis of SZ and X-ray data. The black lines show
    results for a cool-core cluster, the red lines are for the
    isothermal cluster. The distance bias is shown as a function of
    time for an aperture of 1500 kpc ({\it left panel}) and as a
    function of aperture at $t = 7$ Gyr ({\it right panel}).} 
  \label{fig:Dfit_t}
\end{figure*}

We can finally estimate the bias in the determination of the angular 
distance, using equation~(\ref{eq3}). To this end, we again assume
that the observer has performed a de-projection analysis of the X-ray
data and determined the temperature, metal abundance(s) and emission
measure for each spherical shell. We replace the integral in
equation~(\ref{eq3}) by a sum over spherical shells. We find that the
assumption of solar He abundance leads to an underestimate of the
angular distance by $\approx 13$\% and $\approx 15\%$ for the
cool-core and isothermal cluster models respectively, at $t = 7$ Gyr
and using the 1500 kpc aperture. The bias increases almost linearly
with time and weakly depends on the aperture for apertures exceeding
$\sim 250$ kpc in radius. Its detailed dependence on time and
aperture are shown in Fig.~\ref{fig:Dfit_t}.
As one can see from the right panel, in case of A2029 the angular distance bias decreases and even changes sign at small apertures. In particular, there is a value of aperture, about $\sim 20$ kpc in the particular case of A2029, at which the bias equals to zero. This is a result of complex modification in the Helium abundance profiles caused by combined action of gravitational separation and thermal diffusion. However, the position of the zero bias radius depends on the particular shape of the temperature profile. In addition cluster cores are usually difficult to observe because of the complications due to AGN. Therefore decrease of the bias at small apertures and existence of the zero bias aperture is most likely of no practical value for the angular distance measurements.

\section{Summary}

We have considered the role of abundance anomalies in the intra-cluster
medium in interpretating X-ray and microwave observations of
clusters of galaxies in the cosmological context. In particular, we
investigated the role of diffusion of elements in the ICM. As is 
well-known \citep{g84,ch2,ch1,ettori06,sh1}, it is the density
distribution of helium which can be affected by gravitational
sedimentation of elements most strongly. However, since helium  is
fully ionized in the ICM, its abundance (to the contrary to metal
abundances) cannot be directly measured from equivalent widths of
emission lines. Therefore, in practice, it is usually assumed that  
helium abundance in the ICM equals to its primordial value. Should
this assumption be incorrect, significant biases in estimating the
metal abundances, emission measure and total mass of the gas may arise 
\citep[Section 2; see also][]{drake98,ettori06}.  

The role of diffusion in the ICM of clusters of galaxies is still
debated. The two well-known factors that can potentially
significantly reduce its importance are magnetic fields and the mixing
effect of large-scale turbulence. We ignored this complexity and
investigated the maximum possible effect of diffusion on the
cosmological measurements with clusters of galaxies. To this end, we
considered the full set of Burger's equations for a multi-component
plasma and solved them for two cluster models: (i) a cool-core
cluster, which was represented by the temperature and mass profiles of
A2029, and (ii) an isothermal cluster with the same mass distribution
and temperature of $T= 3$ and 6 keV. In the case of the isothermal
cluster, canonical gravitational sedimentation of elements occurs
leading to a factor of $\sim 5$--$10$ enhancement of helium and metal
abundances in the cluster center on a $\sim 3$--$7$ Gyr time
scale. In the case of the cool-core cluster model however, thermal
diffusion counteracts the gravitational sedimentation, significantly
reducing the abundances of all elements heavier than hydrogen in the
cluster inner core, $r\le 10$--$20$ kpc and producing an up to a
factor of $\sim 1.5$--$2$ enhancement of their abundances at the
intermediate radii $\sim 100$--$500$ kpc.   

There is a significant flux of helium and metals and outflow of
hydrogen through the outer boundary of the cluster. This will lead to 
a noticeable increase in the cluster averaged abundances of the elements
heavier than hydrogen at a rate of $\sim 5 \%$  per Gyr. Of course,
for the metals, this depends on their actual abundance in the IGM.  

Such a significant redistribution of elements will lead to a number
of biases in the cluster parameters determined from X-ray data. The
key role in these biases is played by an incorrect assumption about
the helium abundance. In particular, the metal abundances can be
underestimated by up to $\approx 10$--$40\%$ and  the total gas mass
can be underestimated by $\approx 7\%$. When X-ray data are combined
with microwave measurements to measure the angular distance to
clusters of galaxies, the diffusion of elements can lead to an
underestimate of the latter by $\sim 10$--$15\%$ for apertures of
the order $r_{500}$, with the effect being stronger for isothermal
clusters. Furthermore, in the case of an isothermal cluster with
temperature $kT=6$ keV, the effect can reach $\approx 20$--$25\%$ for
small apertures of $\sim 100$--$200$ kpc.  

\section*{ACKNOWLEDGMENTS}

PM and SS acknowledge partial support by the Russian Foundation for Basic Research
(grant 13-02-12250-ofi-m), programs P-21 and OFN-17 of the Presidium of the Russian Academy of Sciences,
and program NSh-6137.2014.2 for support of leading scientific schools in Russia. PM acknowledges the hospitality of Max-Planck Institute for Astrophysics.

\end{document}